\input{aipcheck}

\newcommand {\be} {\begin{equation}}
\newcommand {\ee} {\end{equation}}

\documentclass[
    ,final            
  ]
  {aipproc}

\layoutstyle{6x9}

\begin{document}

\title{Are Quasar Jets Matter or Poynting Flux Dominated?}

\classification{PACS numbers; choose from this list:
               \texttt{http://www.aip.org/pacs/index.html}}

\author{Marek Sikora}{
  address={Nicolaus Copernicus Astronomical Center, Bartycka 18,
00-716 Warsaw, Poland}
}

\author{Greg M. Madejski}{
  address={Stanford Linear Accelerator Center, 2575 Sand Hill Road,
Menlo Park, CA 94025}, altaddress={Kavli Institute for Particle Astrophysics
and Cosmology, Stanford University, Stanford,~CA~94305, US}
}

\author{Jean-Pierre Lasota}{
  address={Institut d'Astrophysique de Paris, UMR 7095 CNRS,
Universit\'e Pierre \& Marie Curie, 98bis,~Boulevard Arago, 75014
Paris, France} }

\author{Mitchell~C.~Begelman}{
address={Joint Institute for Laboratory Astrophysics, University
of Colorado, Boulder, CO 80309-0440, US}
}

\begin{abstract}
If quasar jets are accelerated by magnetic fields but terminate as matter
dominated, where and how does the transition occur between the
Poynting-dominated and matter-dominated regimes?
To address this question, we study constraints which are imposed on the jet
structure by observations at different spatial scales.
We demonstrate that observational data are consistent with a scenario
where the acceleration of a jet occurs within $10^{3-4} R_g$.
In this picture, the non-thermal flares -- important defining
attributes of the blazar phenomenon
-- are produced by strong shocks formed in the region where the jet inertia
becomes dominated by matter. Such shocks may be formed due to collisions
between the portions of a jet accelerated to different velocities,
and the acceleration differentiation is very likely to be related to global
MHD instabilities.
\end{abstract}

\maketitle

\section{1. Introduction}

Extragalactic jets are perhaps the most spectacular products of accretion
activity in quasars and radio galaxies. Yet, despite decades of
observations and intensive theoretical studies, their most fundamental
aspects are still mysterious. It is unclear how they are launched,
accelerated and collimated; why in some active galactic nuclei (AGNs) they
are strong (as a fraction of the total energy output), while in others they
are weak; and whether they are dominated dynamically by matter or magnetic
fields. Various models address these issues, but uncertainties about
the initial and  boundary conditions, as well as the extremely complex
physics of magnetized relativistic outflows, have not allowed a consensus to
be reached concerning the nature of AGN jets.

Recent developments in high-energy astronomy, however, are starting to
provide a way out of this impasse.  X-ray and
$\gamma$-ray observations of blazars, combined with our approximate knowledge
of the central environments in quasars, allow us
to estimate the number and energy flux of electrons/positrons in quasar jets.
The latter is found to be too small to power the observed $\gamma$-ray flares
 or to support the energetics of radio lobes \cite{SM00}.
Therefore, the energy flux in jets must be dominated by protons or magnetic
fields, but with the number of $e^+e^-$-pairs greatly exceeding the number of
protons. We argue in \S2 that production of such jets may
involve mass loading and initial acceleration (in the sub-Alfv\'enic
region) by radiation pressure, and further acceleration by magnetic
stresses.  In \S3, observational constraints on intensity
and structure of magnetic fields in different spatial scales of quasar jets
are discussed. In \S4, we speculate about possible connection of the
blazar activity with the conversion of the  Poynting flux to matter dominated
jets. A more detailed discussion of these issues can be found in \cite{sbml}.

\section{2. Launching a jet}

The most promising scenario for launching quasar jets involves
rotation of large-scale magnetic fields.  The idea of driving
outflows by large-scale magnetic fields, originally proposed
by Weber \& Davis\citep{wd67} to explain the spindown of young
stars, was successfully applied to pulsar winds \cite{Mich69,gj70}
and became a dominant mechanism in theories of relativistic jets in
AGNs \cite{Phin83,Cam86,Cam87,Love87,lcb92,vk04}. Powerful,
magnetically dominated outflows can be driven from both an accretion
disk and a black hole  magnetosphere. Such outflows can  become
relativistic if the total to rest-mass energy flux ratio $\mu \equiv
L_j / \dot M c^2 \gg 1$, where $L_j= L_B + L_{kin}$ is the total
energy flux, $L_B$ is the magnetic energy flux,
$L_{kin}=(\Gamma-1)\dot M c^2$ is the kinetic energy flux, and $\dot
M$ is the mass loading rate. Following the work by Blandford \&
Payne\citep{bp82}, it is often claimed that without sufficient
thermal pressure the MHD outflows from the disk can be produced only
for magnetic field lines inclined at $i>30$ degrees to the disk
rotation axis. For such angles the effective potential is decreasing
along magnetic field lines  and the outflow can be launched and
driven away by centrifugal forces even for very low coronal
temperatures. This may lead to very efficient  mass loading and,
therefore, to non-relativistic terminal velocities
\cite{ops97,ukr99}.
 For $i < 30$ degrees, the coronal plasma cannot be
freely driven by centrifugal forces:  instead, it must first
overcome the effective-potential barrier, which can have its
maximum far away from the disk. Therefore, a strong thermal or
radiative assistance is  required to initiate outflows
in such a geometry.
The latter can be particularly efficient in quasars that radiate at
a significant  fraction of the Eddington rate and have pair rich coronae.
Preliminary studies of launching and developing outflows with $\mu \gg 1$
have been recently performed both analytically and numerically
\cite{LB96,meg97,llfc01,ulr00,llkur02,lr03}, but none of these works
addressed the issue of mass loading and acceleration in the
sub-Alfv\'enic region for typical quasar conditions.

A basic question regarding scenarios for the formation of powerful,
relativistic MHD jets by accretion disks concerns the origin of the
strong poloidal magnetic field.  Two possibilities have been
considered in the literature: one is that such magnetic fields are
advected inward from the interstellar medium by accreting matter
\cite{bp82}, and the other one is that they are generated locally by
a dynamo \cite{rbds03}. The first one is often questioned because
the dragging of magnetic fields inward requires the magnetic Prandtl
number to be unrealistically large \cite{lbp94,hpb96}. However, this
argument applies only if the accretion is driven by viscous 
torques in the turbulent disk. If the angular momentum is carried
away by the MHD wind, then the magnetorotational instability (MRI)
that drives the turbulence is suppressed and magnetic field
advection can become efficient. The second possibility is often
criticized because the large-scale fields produced by the dynamo are
expected to be predominantly toroidal \cite{tbr99,ms00,kms02}. Under
certain circumstances, however, an inverse cascade of reconnecting
magnetic loops could produce a dominant poloidal component
\cite{tp96,lpk03} and accretion could then be driven by the torque
exerted on the disk by the MHD outflow.

The production of very strong and relativistic jets requires a large fraction
of the gravitational energy of accreting matter to be converted to Poynting
flux.  This condition can be satisfied only in the very central region, but
the collimation of such a jet requires the disk to be threaded by a poloidal
 magnetic field over much larger scales \cite{sfs97}.
The collimation/confinement of central, weakly mass-loaded,
electromagnetic outflows is then provided by slower and more massive
MHD outflows, launched at larger disk radii by centrifugal forces
\cite{tb02,bt05}. One particular version of such a hybrid  outflow
model has been suggested by Sol, Pelletier \& Asseo\citep{spa89}.

\section{3. Magnetically Dominated Over Which Scales?}

If a jet is launched magnetically, does it remain magnetically dominated over
all scales up to the termination shock, or does it undergo conversion to
a kinetic energy-dominated state?  The theory of axisymmetric, steady-state
ideal MHD outflows predicts that the conversion process works efficiently
up to the classical fast-magnetosonic surface, $z_f$,  which is  located
at a few light cylinder-radii \cite{Sak85,lcb92,bkr98}. At this distance, 
the ratio of
Poynting flux to kinetic energy flux, $\sigma$, drops  to the value
$\sim \mu^{2/3}$. This means that for $\mu \gg 1$ the flow still remains
strongly Poynting flux-dominated at $z_f$. Whether and how fast the conversion
 can proceed beyond this point depends on
the very uncertain boundary conditions \cite{bl94,hn03,vk04,bn05}.
Below, we discuss whether there is any observational evidence of the dynamical
dominance of magnetic fields on any scale in quasar jets.

\subsection{3.1. The blazar zone}

Blazar variability timescales of $\sim 1$ week in the optical band and
similar or even shorter fluctuations with larger amplitudes in the
$\gamma$-ray band \cite{mbc95,mbb97} show
that most of non-thermal radiation in quasar jets is produced within
a few parsecs from the center. This is
independently confirmed by the location of the cooling break in blazar
spectra \cite{msb03}. Polarization of the variable optical, infrared
and mm radiation suggests
the dominance of perpendicular magnetic fields in the blazar jets
\cite{ilt91,gs94,cg96,srh96,ngm98}. Such an orientation
is consistent with a toroidal magnetic field geometry, but can also result
from compression of a tangled magnetic field in a transverse shock.
Such shocks have been proposed to result from collisions between velocity
inhomogeneities propagating down a matter-dominated jet  \cite{SBR94,sglc01}.
This internal shock scenario is
supported by the very broad energy distributions of relativistic
electrons/positrons. They cover 3-4 decades in energy and are injected with
approximately equal amounts of energy per decade \cite{msb03}.
This contrasts strongly with the narrow energy distributions of accelerated
electrons predicted by the magnetic reconnection models \cite{ZH01,LLR03}.

\subsection{3.2. Parsec scales}

There are phenomenological arguments in favor of the dynamical domination of
 magnetic fields in parsec-scale jets.  Some of these arguments are based on
VLBI observations of the superluminal propagation of radio features.
If such features were carried by a Poynting flux-dominated jet, they
should be accelerating.   Homan et al.\citep{how01} claim that in
sources having multiple components with measurable proper motion,
the innermost components are significantly slower than the others.
If true, this would suggest that indeed the flow is accelerating.
However, the assertion about slower moving innermost components seem
to contradict the finding that there is a systematic decrease in
apparent velocity with increasing wavelength \cite{klh04}. The
simplest interpretation of this is that the observations at longer
wavelengths cover more extended portions of the jet structure, and
therefore that the radio components decelerate, rather than
accelerate. Noting also that some outflows bend or change 
their opening angle, one should not be surprised to see both
increasing and decreasing projected speeds.  In these cases, one
learns little about the intrinsic kinematics of the source from the
motion of the surface-brightness-peak of the radio component.  This
is because such peaks probably do not represent the real component
centers, due to the relativistic aberration and Doppler effects from
intrinsically expanding finite-size sources. Furthermore, even if some 
apparent acceleration events
are real, they are not necessarily related to the conversion of
magnetic energy to kinetic energy. Acceleration events can be
produced also in matter-dominated jets, e.g., at the expense of
energy dissipated in shocks and partially
returned to the flow, or  can be represented
by shocks formed on the interface between a jet and a clump of matter
entering the jet from outside and being accelerated by the relativistic flow.
Finally, the features that appear as moving on the VLBI scale may represent
moving patterns rather than the real flow speeds.
Noting all the above,  we would consider as premature claims that
``accelerating'' individual features in 3C 279 \cite{puw03} and
3C 345 \cite{uwl97,lz99}
indicate magnetic domination of parsec-scale jets in these objects.

Another approach to studying the dynamics of a jet is based on
comparing its surface brightness distribution with that of its
counterjet. This method was applied by Sudou et
al.\citep{so02} to prove acceleration of a jet in NGC 6251; however,
the reality of the counter-jet detection in this object is
questioned  by Jones \& Wehrle\citep{jw02}.  Furthermore, it should
be emphasized that this method is based on the assumption that the
jet is steady, whereas parsec scale jets  are usually variable. For
unsteady jets, even if they are intrinsically symmetric, the
respective flux ratios are expected to vary  due to
light--travel--time effects and, therefore, multiple  observing
campaigns are needed to verify any premises about the flow
acceleration.

The presence of strong, ordered  magnetic fields in jets could
eventually be diagnosed  by studies  of gradients of the rotation
measure (RM) across a jet. Using this method, Gabuzda, Murray \&
Cronin\citep{gmc03} found evidence for toroidal field in several BL
Lac objects. The RM gradient was found also in quasar 3C 273
\cite{aiu02,zt05,awh05}. However, the fact that Faraday rotation in
many objects follow the '$\lambda^2$-rule', even in objects with
rotation exceeding 1 radian imply its external origin. On another
hand, the time variability \citep{zt05} and the rapid decrease of
the RM gradient with distance down the jet \cite{awh05} indicate
that Farady screen is located very nearby the jet. The screen can be
provided by slower moving outer portions of the structured jet.

The presence of the toroidal magnetic component in quasar jets is
indicated by measurements of the circular polarization
\cite{whor98,hw99,haw01}. However, as
was demonstrated by Ruszkowski \& Begelman\citep{rb01}, the observed
circular polarization features can be explained without invoking
strong, ordered magnetic fields.

\subsection{3.3. Kiloparsec scales}

Often-used arguments in favor of the dynamical dominance of magnetic fields
over large spatial scales include the high linear polarization of
kiloparsec-scale jets, and the need for ``in situ'' energy dissipation
to provide fast-cooling ultra-relativistic electrons responsible
for synchrotron radiation in the optical and X-ray band  \cite{lb98,bl00}.
However, high polarization does not
necessary require large scale mean magnetic fields;
it can be produced in shocks and in  boundary shear layers \cite{so02},
where initially tangled/turbulent magnetic fields
are ordered by compression and stretching, respectively
\cite{Lai80,Lai81,cwrg93}.
 The parallel magnetic field orientation indicated by polarimetry of
large-scale radio jets in FRII radio galaxies and quasars
\cite{bhlbl94} suggests that shear layers play a dominant role in
powering the emission from large-scale jets.  Direct support for
this scenario is provided by measurements of intensity and
polarization profiles across jets in a number of nearby objects
\cite{sbb98}. The perpendicular electric vector orientation in
respect to the jet axis can result also from compression of tangled
magnetic fields by oblique shocks. This can explain the
perpendicular polarization of optical light in 3C 273 jet
\cite{rm91,tmw93}. Since formation of strong oblique shocks is
unlikely to take place in the presence of a magnetically dominated
jet, confirmation of the perpendicular orientation of the electric
vector in the optical band in 3C 273 and other quasars can prove
that in kilparsec scale jets, $\sigma \ll 1$.

The hydrodynamical nature of the large scale jets is also indicated
by numerical simulations of their propagation. As Clarke, Norman, \&
Burns\citep{cnb86} and Lind et al.\citep{lpmb89} demonstrated for
non-relativistic jets and Komissarov\citep{Kom99} showed for
relativistic jets, magnetically dominated jets do not develop
substantial back-flowing cocoons. Instead, the shocked jet plasma,
being confined by magnetic stresses, forms a ``nose cone'' -- shaped
head. The cocoons observed in classical FR II radio sources do not
form such nose-cones. They are broad and their morphologies agree
very well with the cocoons predicted by numerical simulations of
light, supersonic, unmagnetized jets.  Although there are a few
radio quasars that possess a nose-cone radio morphology, this by
itself does not prove the dominance by magnetic fields. As
Komissarov \& Falle\citep{KF98} pointed out, a nose-cone morphology
can also result if a jet is heavy, i.e., if its co-moving density
multiplied by the Lorentz factor is larger than the density of the
ambient plasma. This condition
 can be satisfied, for example, if the source is intermittent and the jet is
restarted into the old, expanded cocoon --- the remnant of an earlier epoch
 of activity.  Stawarz\citep{Sta04b} has proposed such an interpretation for
the unusual morphology of 3C 273.

\section{4. Conclusions}

Quasar jets are presumably launched by rotating magnetic fields
in the vicinity of super-massive black holes and, as MHD theories
predict and bulk-Compton constraints support,  are
magnetically dominated  over at least three distance decades.
There appears to be no evidence of magnetic field domination on parsec and
larger scales, and this
suggests that the conversion of a magnetically-dominated
to matter dominated jet takes place within  the blazar zone.
Such a location of the conversion is independently supported by data on
kinematics of a jet. Radiation models of high energy flares in blazars
give a bulk Lorentz factor  $\Gamma  \sim 10 - 20$ \cite{gcf98}.
Lorentz factors of the same order
are directly monitored by radio interferometers on parsec scales
\cite{how01,jmm01,klh04}
and inferred from X-ray and optical observations on kiloparsce scales
\cite{tmsu00,tmsu04,SU02}. On the other hand,
the lack of signatures of bulk-Compton radiation in the blazar spectra
implies much slower flows prior to the blazar zone
\cite{SM00,msmk04}.  It is tempting to speculate that short term,
high amplitude flares in blazars are related to MHD instabilities
\cite{Eich93,beg98}, developed in a jet during its final stages of
acceleration.

\begin{theacknowledgments}
This project was supported by the Polish grant PBZ-KBN-054/P03/2001.
\end{theacknowledgments}

\end{document}